\documentclass[11pt]{article}
\usepackage{amsfonts,amscd,amsmath, amssymb,graphicx,color}
\def\pot{V_{\text{\tiny 3q}}}
\def\potS{\tilde V_{\text{\tiny 3q}}}
\def\tetra{V_{\text{\tiny 4q}}}
\def\alp{\alpha_{\text{\tiny 3q}}}
\def\sig{\sigma_{\text{\tiny 3q}}}
\def\sigN{\sigma_{\text{\tiny Nq}}}
\def\sigS{\tilde\sigma_{\text{\tiny 3q}}}
\def\lmin{L_{\text{\tiny min}}}
\def\W{W_{\text{\tiny 3q}}}
\def\WS{\tilde W_{\text{\tiny 3q}}}
\def\tens{\sigma_{\text{\tiny q\={q}}}}
\def\alpq{\alpha_{\text{\tiny q\={q}}}}
\def\f0{f_{\text{\tiny 0}}}
\newcommand{\oh}{\frac{1}{2}}
\def\z{z_{\text{\tiny 0}}}
\def\ep{\text{e}}
\def\rt{r_{\text{\tiny T}}}
\def\T{T_c}

\def\g{\mathfrak{g}}
\def\oh{\frac{1}{2}}
\def\x0{x_{\text{\tiny 0}}}
\def\y0{y_{\text{\tiny 0}}}
\def\r0{r_{\text{\tiny 0}}}
\def\z0{z_{\text{\tiny 0}}}
\def\m{\mathfrak{m}}
\def\s{\mathfrak{s}}
\def\last{\lambda_\ast}
\def\h{\text{e}^{\s r^2}}

\title{Some Multi-Quark Potentials, Pseudo-Potentials and AdS/QCD}
\author{Oleg Andreev \thanks{Also at Landau Institute for Theoretical Physics, Moscow. }\\\\
{\it Technische Universit\"at M\"unchen, Excellence Cluster,} \\
{\it Boltzmannstrasse 2, 85748 Garching, Germany}
}
\date{}

\begin{document}

\maketitle
\begin{abstract}
The static three-quark potential and pseudo-potential of a pure $SU(3)$ gauge theory are studied in a five-dimensional framework known as 
AdS/QCD. The results support the Y-ansatz for the baryonic area law. A comparison with the quark-antiquark calculations shows 
the universality of the string tension as well as the spatial string tension. We also discuss extensions to $SU(N)$ gauge theories. 
 \end{abstract}

\vspace{-13 cm}
\begin{flushright}

{\small SPAG-A1/08}\\
\end{flushright}

\vspace{13 cm}


\section{Introduction}
\renewcommand{\theequation}{1.\arabic{equation}}
\setcounter{equation}{0}
Heavy-quark potentials are of primary importance in studying the mass spectra of mesons and baryons. They have been computed in 
lattice simulations, and the results reveal a remarkable agreement with phenomenology.\footnote{For a review, see \cite{rev-lat}.}

Until recently, the lattice formulation even struggling with limitations and systematic errors was the main computational tool to deal with 
strongly coupled gauge theories. The situation changed drastically with the invention of the AdS/CFT correspondence that resumed interest 
in finding a string description of strong interactions.

In this paper we continue a series of studies \cite{az1,az2,andreev} devoted to the heavy quark potentials and pseudo-potentials within a 
five-dimensional framework nowadays known as AdS/QCD. In \cite{az1}, the model was presented for computing 
the quark-antiquark potential. The resulting potential is Coulomb-like at short distances and linear at long 
range. Subsequent work \cite{list} has made it clear that the model should be taken seriously, particularly in the context of consistency with the 
available lattice data as well as phenomenology. In \cite{az2,andreev}, the models were  presented for computing the quark-antiquark 
pseudo-potentials resulting from the spatial Wilson loops. The results obtained for the spatial string tensions are remarkably consistent with the 
available lattice data for temperatures up to $3\T$.

The question naturally arises: What happens when these models are used for computing multi-quark potentials or pseudo-potentials? The 
multi-quark potentials have recently been the object of numerical studies.\footnote{For reviews, see \cite{lattice,lattice2} and references therein.} In 
the case of great interest, three quark states in a $SU(3)$ gauge theory, the potential is well described by a simple model \cite{3q-lattice}

\begin{equation}\label{lat-potential}
\pot =-\alp\sum_{i<j}^3\frac{1}{L_{ij}}
+\sig \,\lmin+C
\,.
\end{equation}
Here $C$ is a constant. The quarks form a triangle with sides of lengths $L_{ij}$. $\lmin$ is the minimal length of the string network which 
has a junction at the Fermat point of the triangle \cite{fermat}. Thus, the potential is given by the sum of the Coulomb terms and the 
linear term called the Y-law. A remarkable fact is that $\sig$ is equal to $\tens$ found from the quark-antiquark potential. This is obvious in 
the string picture, where the string tension is universal, but it is far from being so in the lattice formulation. 

The purpose of the present paper is two-fold. First, we examine the multi-quark potentials that may also be thought of as a further cross-check of the 
model \cite{az1}. Second, we make a similar analysis of the multi-quark pseudo-potentials. To our knowledge, there have been no studies 
(numerical or analytical) of this problem in the literature. 

The paper is organized as follows. In section 2, we discuss the multi-quark potentials. We begin with the three-quark potential. In this case, we 
demonstrate the Y-law and the universality of the string tension. Finally, we extend our analysis to the $SU(N)$ case. We then go on in section 3 to 
discuss the multi-quark pseudo-potentials. Here we also demonstrate the Y-law, the universality of the spatial string tension, and possible 
generalizations to $SU(N)$. We conclude in section 4 with a brief discussion of possibilities for further study. Some technical details are given in 
the appendix.
 
\section{Calculating the Potentials }
\renewcommand{\theequation}{2.\arabic{equation}}
\setcounter{equation}{0}

In this section, we will discuss static multi-quark potentials. We start with the three-quark potential in a $SU(3)$ gauge theory, where 
equations are most elementary.  In section 2.4, similar issues are considered in the context of a $SU(N)$ gauge theory. 

\subsection{General Formalism}

As for the quark-antiquark potential, the static three-quark potential can be determined from the expectation value of  a Wilson loop. The baryonic 
loop is defined in a gauge-invariant manner as $\W=\frac{1}{3!}\varepsilon_{abc}\varepsilon_{a'b'c'} U_1^{aa'}U_2^{bb'}U_3^{cc'}$, 
with the path-ordered exponents $U_i$ along the lines shown in Figure 1.

%
\begin{figure}[ht]
\begin{center}
\includegraphics[width=2.5cm]{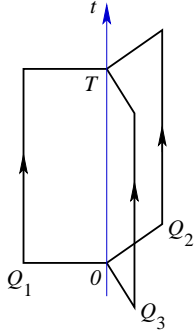}
\caption{\small{A baryonic Wilson loop $\W$. A three-quark state is generated at $t=0$ and is annihilated at $t=T$. The quarks are spatially
fixed in $\text{\bf R}^3$ at points $Q_1$, $Q_2$ and $Q_3$. }}
\end{center}
\end{figure}

In the limit $T\rightarrow\infty$ the expectation value of the Wilson loop is 

\begin{equation}\label{wloop}
\langle\,\W \,\rangle\sim \ep^{-T\pot}
\,,
\end{equation}
where $\pot$ is the three-quark potential.

In discussing baryonic Wilson loops, we adapt a formalism proposed within the AdS/CFT correspondence \cite{witten, gross} to 
AdS/QCD.\footnote{For subsequent developments of the formalism, see \cite{son,liu}.} So, we place heavy quarks at the 
boundary points of the five-dimensional space and consider a configuration in which each of the quarks is the endpoint of a fundamental 
string, with all the strings oriented in the same way. The strings join at a baryon vertex in the interior as shown in Figure 2.\footnote{From the 
point of view of 10-dimensional string theory, the baryon vertex is a wrapped fivebrane whose world-volume is ${\bf R}\times {\bf X}$, 
with ${\bf X}$ a 5-dimensional compact space and ${\bf R}$ a {\it timelike} curve in $\text{AdS}_5$ or its deformed version, \cite{witten, gross}. 
In AdS/QCD, it is reduced to a zerobrane (point). } 
\begin{figure}[ht]
\begin{center}
\includegraphics[width=4.75cm]{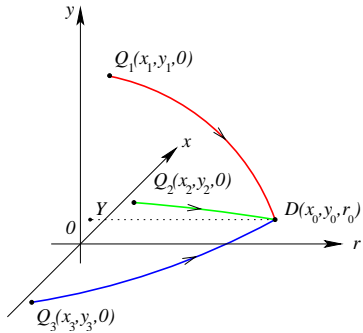}
\caption{\small{A configuration used to calculate the expectation value of $\W$. The quarks are set on the $x\text{-}y$ plane. The 
baryon vertex is placed at $D$. Its projection on the $x\text{-}y$ plane is $Y$.}}
\end{center}
\end{figure}
We also assume that the quarks form a triangle $Q_1Q_2Q_3$ such that all the internal angles are smaller than $\tfrac{2\pi}{3}$.

Before proceeding to the detailed analysis, let us set the five-dimensional geometry. We consider the following deformation of the Euclidean 
$\text{AdS}_5$ \cite{oa,az1}

\begin{equation}\label{metric}
ds^2=R^2 w \bigl(dt^2+d\vec x^2+dr^2\bigr)\,,
\quad w(r)=\frac{\h}{r^2}
\,,
\end{equation}
where $d\vec x^2=dx^2+dy^2+dz^2$ and $\s$ is a free parameter.\footnote{The value of $\s$ can be fixed, for example, from 
the quark-antiquark potential.} We also take a constant dilaton and discard other background fields. 

The action of the system has in addition to the standard Nambu-Goto actions of the fundamental strings, also a contribution arising from the 
baryon vertex.  It is thus 

\begin{equation}\label{stringaction}
{\cal S}=\sum_{i=1}^3  S_i +S_{\text{\tiny vert}}
\,,
\end{equation}
where $S_i$ denotes the action of the string connecting the $i$-quark with the vertex.

Like in the case of the quark-antiquark potential, a natural proposal for the expectation value of the Wilson loop is

\begin{equation}\label{wloop-ads}
\langle\,\W \,\rangle\sim \ep^{-{\cal S}_{\text{\tiny min}}}
\,,
\end{equation}
where ${\cal S}_{\text{\tiny min}}$ is the minimal action of the system. 

Since we are interested in a static configuration, we take 
 
\begin{equation}\label{gauge}
t_i(\tau_i)=\tau_i\,,\quad
y_i(\sigma_i)=a_i\sigma_i+b_i
\,.
\end{equation}
Here $(\tau_i,\sigma_i)$ are worldsheet coordinates. The action of the $i$-string is then
\begin{equation}\label{ng-action}
S_i=T\g\int_0^1 d\sigma_i\,w\sqrt{a^2_i+ x'^2_i+r'^2_i}
\,,
\end{equation}
where $\g=\tfrac{R^2}{2\pi\alpha'}$. A prime denotes a derivative with respect to $\sigma_i$.

The action for the baryon vertex is taken to be of the form

\begin{equation}\label{vertex}
S_{\text{\tiny vert.}}=T{\cal V}(\r0 )
\,,
\end{equation}
where ${\cal V}$ can be considered as an effective potential of the vertex. Unfortunately, the explicit form of ${\cal V}$ is not determined only 
from the 5-dimensional metric \eqref{metric}. It requires the knowledge of the string theory dual to QCD. We will return to this issue in the 
next section.

The boundary conditions on the fields are given by 
\begin{equation}\label{boundary}
x_i(0)=x_i\,,
\quad
y_i(0)=y_i\,,
\quad
r_i(0)=0\,,
\quad
x_i(1)=\x0\,,
\quad
y_i(1)=\y0\,,
\quad
r_i(1)=\r0\,.
\end{equation}
These determine the coefficients $a_i$  and $b_i$ in \eqref{gauge}. Thus, we have

\begin{equation}\label{y}
a_i=\y0 -y_i\,,
\qquad
b_i=y_i
\,.
\end{equation}

Next, we extremize the total action $\cal S$ with respect to the worldsheet fields $x_i(\sigma_i)$ and 
$r_i(\sigma_i)$ describing the strings as well as with respect to $\x0$, $\y0$ and $\r0$ describing the location of the baryon vertex, with the 
following identifications: $\delta x_i(1)=\delta\x0$ and $\delta r_i(1)=\delta\r0$. In doing so, we use the fact that there are two symmetries 
which simplify the further analysis. 

Since the integrand in \eqref{ng-action} does not depend explicitly on $\sigma_i$, we get the first integral of Euler-Lagrange equations

\begin{equation}\label{1integral}
I_i=\frac{w_i}{\sqrt{a_i^2+x_i'^2+r_i'^2}}
\,.
\end{equation}
In addition, because of translational invariance along the $x$-direction, there is another first integral. Combining 
it with \eqref{1integral} gives

\begin{equation}\label{2integral}
P_i=x'_i\,.
\end{equation}
Together with the boundary conditions these equations determine $x_i$

\begin{equation}\label{x}
x_i (\sigma_i )=(\x0 -x_i)\sigma_i +x_i
\,.
\end{equation}

Now, we extremize the action with respect to the location of the baryon vertex. After using \eqref{y} and \eqref{x}, we get 

\begin{equation}\label{xyr-eqs}
\sum_{i=1}^3 \frac{\x0 -x_i}{l_i\sqrt{1+k_i}}=0
\,,
\qquad
\sum_{i=1}^3 \frac{\y0 -y_i}{l_i\sqrt{1+k_i}}=0
\,,
\qquad
\sum_{i=1}^3\frac{1}{\sqrt{1+k_i^{-1}}}+
\frac{1}{\g}\frac{{\cal V}'}{w}(\r0)=0
\,.
\end{equation}
Here $k_i=\left(\frac{r'_i(0)}{l_i}\right)^2$ and $l_i=\vert Y Q_i\vert=\sqrt{(\x0 -x_i)^2+(\y0 -y_i)^2}$.
 
If we define the first integrals $I_i$ at $\sigma_i=1$ such that $I_i=\omega(\r0 )/l_i\sqrt{1+k_i}$, and then integrate over $[0,1]$ of 
$d\sigma_i$, then by virtue of \eqref{1integral} we get

\begin{equation}\label{l-eqs}
l_i=
\sqrt{\frac{\lambda}{\s(1+k_i)}}
\int_0^1 dv_i\,v_i^2\ep^{\lambda(1-v_i^2)}
\biggl(1-\frac{1}{1+k_i}v_i^4\ep^{2\lambda(1-v_i^2)}\biggr)^{-\oh}
\,,
\end{equation}
where $v_i=r_i/\r0$ and $\lambda=\s\r0^2$.

Now, we will compute the energy of the configuration. First, we reduce the integrals over $\sigma_i$ in Eq.\eqref{ng-action} to that 
over $r_i$. This is easily done by using the first integral \eqref{1integral}. Since the integral is divergent at $r_i=0$, we regularize it by imposing 
a  cutoff $\epsilon$. Then we replace $r_i$ with $v_i$ as in \eqref{l-eqs}. Finally, the regularized expression takes the form
\begin{equation}\label{energy}
E_R=
{\cal V}(\lambda)+
\g\sqrt{\frac{\s}{\lambda}}\sum_{i=1}^3
\int_{\sqrt{\frac{\lambda}{\s}}\epsilon}^1 \frac{dv_i}{v_i^2}\, \ep^{\lambda v_i^2}
\biggl(1-\frac{1}{1+k_i}v_i^4\ep^{2\lambda(1-v_i^2)}\biggr)^{-\oh}
\,.
\end{equation}
Its $\epsilon$-expansion is 
\begin{equation*}\label{energy2}
E_R=\frac{3\g}{\epsilon}+O(1)
\,.
\end{equation*}
Subtracting the $\frac{1}{\epsilon}$ term (quark masses) and letting $\epsilon=0$, we find a finite result

\begin{equation}\label{energy3}
E={\cal V}(\lambda)
+\g\sqrt{\frac{\s}{\lambda}}\sum_{i=1}^3
\int_0^1 \frac{dv_i}{v_i^2}\Biggl[\ep^{\lambda v_i^2}
\biggl(1-\frac{1}{1+k_i}v_i^4\ep^{2\lambda (1-v_i^2)}\biggr)^{-\oh}-1-v_i^2\Biggr] + C
\,,
\end{equation}
where $C$ stands for a normalization constant. 

In contrast to the quark-antiquark case, the potential in question is more involved. It is given by a set of equations. Formally, one can eliminate 
parameters and find $E$ as a function of the $l_i$'s or the $L_{ij}$'s. Unfortunately, in practice it is extremely difficult. 

\subsection{A Concrete Example}
We will next describe a concrete example in which one can develop a level of understanding that is somewhat similar to that of the 
quark-antiquark case \cite{az1}. We consider the most symmetric configuration of the quarks, in which the triangle $Q_1Q_2Q_3$ is 
equilateral, and specify the action $S_{\text{\tiny vert}}$ as that of a particle in a curved space. The latter implies that   

\begin{equation}\label{potential}
{\cal V}(\r0 )=\m R\sqrt{\omega (\r0)}
\,,
\end{equation}
where $\m$ is a parameter, which can be interpreted as a mass of a "particle''.

A consequence of the symmetry is that $Y$, which is the projection of $D$ on the $x$-$y$ plane, is nothing but the circumcenter of the 
triangle $Q_1Q_2Q_3$. The first two equations of \eqref{xyr-eqs} are now identically satisfied, while the last takes the form\footnote{This 
equation determines the location of the vertex in the $r$-direction, because the symmetry argument alone is not enough to do so.}

\begin{equation}\label{r-eqs}
\frac{1}{\sqrt{1+k^{-1}}}=\kappa\left(1-\lambda\right) \ep^{-\oh\lambda}
\,,
\end{equation}
where $k=k_i$ and $\kappa=\tfrac{1}{3}\frac{\m R}{\g}$. 

With the form \eqref{r-eqs}, $k$ can be explicitly computed as a function of $\lambda$. Inserted in \eqref{l-eqs} and \eqref{energy3}, this gives

\begin{equation}\label{l-eqs-ex}
l=\sqrt{\frac{\lambda}{\s}\rho}\int_0^1 dv\,v^2\ep^{\lambda (1-v^2)}\Bigl(1-\rho \,v^4\ep^{2\lambda (1-v^2)}\Bigr)^{-\oh}
\,
\end{equation}
and
\begin{equation}\label{energy-ex}
E=3\g\sqrt{\frac{\s}{\lambda}}
\biggl[\kappa\,\ep^{\oh\lambda}+
\int_0^1 \frac{dv}{v^2}\biggl(\ep^{\lambda v^2}
\Bigl(1-\rho \,v^4\ep^{2\lambda (1-v^2)}\Bigr)^{-\oh}-1-v^2\biggr)\biggr]+C
\,,
\end{equation}
where $\rho(\lambda)=1-\kappa^2\left(1-\lambda\right)^2 \ep^{-\lambda}$.

The potential in question is written in parametric form given by Eqs.\eqref{l-eqs-ex} 
and \eqref{energy-ex}.\footnote{Note that these equations are reduced to those of \cite{az1} at $\kappa=0$ and $\rho=1$.} It is unclear to us 
how to eliminate the parameter $\lambda$ and find $E$ as a function of $l$. We can, however, gain some important insights into the problem 
from two limiting cases as well as numerical calculations.

We start by noting that Eq.\eqref{r-eqs} makes sense only if $\lambda<1$. This means that $\r0$ must obey 
\begin{equation}\label{wall}
\r0\leq \sqrt{\frac{1}{\s}}
\,.
\end{equation}
So in other words, the baryon vertex is prevented from getting deeper into the $r$-direction. This gives a kind of wall which is a generic feature of 
confining theories. It is worth mentioning that the same upper bound was found in the quark-antiquark case \cite{az1} by inspecting the 
integral \eqref{l-eqs-ex} at $\rho =1$. The point is that the integral is real for $\lambda <1$. It develops a logarithmic 
singularity at $\lambda=1$ and becomes complex for larger $\lambda$. A similar analysis shows that if $\kappa\leq 1$ this holds for 
smaller $\rho$ too. For $\kappa>1$ things are more subtle because of a lower bound 
\begin{equation}\label{wall2}
r_\ast\leq\r0
\,.
\end{equation}
Here $r_\ast$ is a root of $\rho(\s r^2)=0$. The physical reason for this is a big mass of the "particle'' that strengthens a gravitational force 
pushing the baryon vertex deeper into the interior. As a result, for small $l$ the configuration looks like a spike of height $r_\ast$. 

The fact that both the equations \eqref{r-eqs} and \eqref{l-eqs-ex} lead to the same upper bound suggests that neither strings nor baryon vertices are 
allowed to get deeper into the $r$-direction than $1/\sqrt{\s}$. This seems natural enough from the point of view of consistency.

To complete the picture, let us present the results of numerical calculations. The parametric equation \eqref{l-eqs-ex} predicts a characteristic 
form of $l$, as shown in Figure 3. 
\begin{figure}[ht]
\begin{center}
\includegraphics[width=5.25cm]{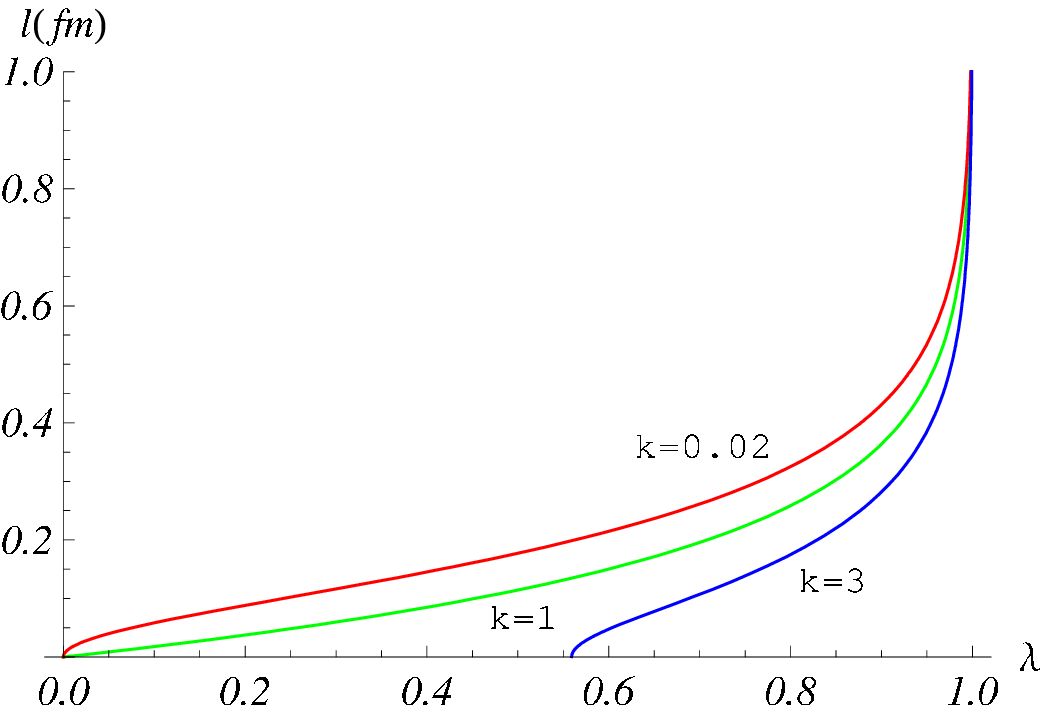}
\caption{\small{Plots of $l(\lambda,\kappa)$ for fixed $\kappa$. Here $\s =0.45\,\text{GeV}^2$.}}
\end{center}
\end{figure}
We see that the curves behave similarly in the vicinity of $\lambda=1$ but rather differently for smaller $\lambda$. This has an interesting 
effect on the form of the interquark potential, as we will see in a moment.

Having understood the correspondence between $\lambda$ and $l$, we can investigate the properties of the interquark interaction at long and 
short distances.\footnote{More details are presented in the appendix.}

At long distances the interquark interaction being independent of $\kappa$ is given by 

\begin{equation}\label{Y-law1}
E=\sig\lmin +O(1)
\,
\end{equation}
that is nothing but the desired Y-law. Here $\lmin=3l$ and $\sig=\ep\g\s$.

Some comments about formula \eqref{Y-law1} are in order. First, the form of the potential is in agreement with both the old flux-tube picture of 
hadrons and more recent calculations in lattice QCD \cite{lattice,lattice2}. However, the key difference is that in our model the interaction 
takes place in the interior of five-dimensional space, whereas in those it happens on its boundary. Next, the tension $\sig$ is the same as that 
found from the quark-antiquark potential \cite{az1}. Thus, the model also supports the universality of the string tension. Finally, to leading order the 
energy is the sum of the string contributions. A contribution from the baryon vertex appears at next-to-leading-order. 

At short distances the form of the interquark interaction depends on the value of $\kappa$. We have the Coulomb terms

\begin{equation}\label{Coulomb1}
E=-3\frac{\alp}{L}+O(1)
\,,
\end{equation}
with $L$ a distance between the quarks, if and only if $\kappa<1$. In other cases, the energy behaves as $E\sim\text{const}$. 

Just as in the examples discussed in the AdS/CFT formulation \cite{witten, son}, in \eqref{Coulomb1} the contribution of the baryon vertex is 
compatible to that of the strings. The reason for this is obvious: in the region of small $r$ the metric \eqref{metric} behaves asymptotically as 
Euclidean $\text{AdS}_5$.

As an illustration, Figure 4 shows the results of numerical calculations. We see that the value of $\kappa$ does matter
\begin{figure}[ht]
\begin{center}
\includegraphics[width=5.25cm]{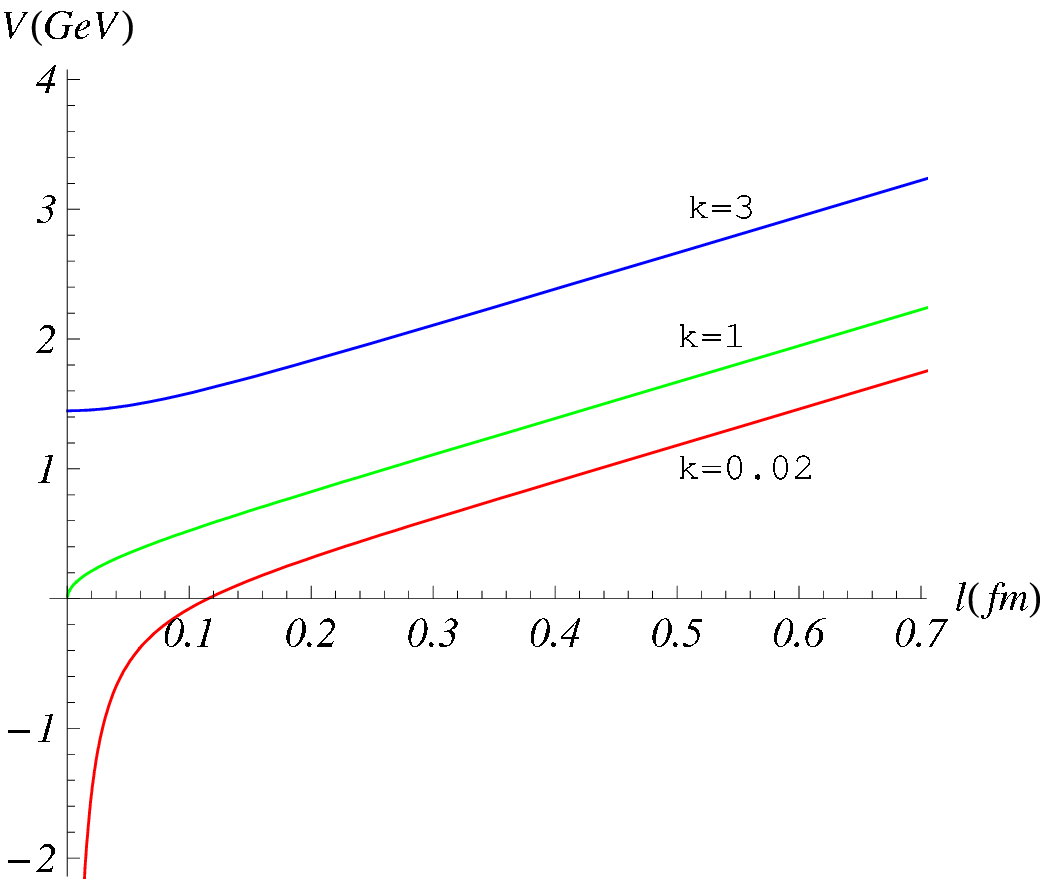}
\caption{{\small Plots of $E(l,\kappa)$ for fixed $\kappa$. We set $\g=0.15$, $\s =0.45\,\text{GeV}^2$, $C=0$.}}
\end{center}
\end{figure}
at short distances, while it becomes irrelevant at large distances. 

We conclude this section by making a few estimates, which might be of interest for phenomenology.

As already noted, in AdS/CFT the baryon vertex is the wrapped fivebrane. For definiteness, consider type IIB string theory on 
$AdS_5\times S^5$. The D5 brane world-volume is then $S^5\times R$, with $R$ a timelike curve in $AdS_5$. Now, suppose that 
the action of the brane is simply the Nambu-Goto term ${\tau_5}\int d^6\xi \sqrt{g}$, with ${\tau_5}$ the tension.\footnote{The general 
form of the action is given by the DBI and WZ terms plus $\alpha'$ corrections. See, e.g., \cite{joe}.} In the static case this leads to the same form 
as \eqref{potential}. Moreover, we learn that $\kappa=\tfrac{1}{4}$. 

At first sight, it seems reasonable to fix the value of $\kappa$ by equating $\alp$ and $\oh\alpq$.\footnote{With accuracy better 
than a few percent, such a relation is valid in lattice QCD \cite{3q-lattice}.} A little experimentation with Mathematica shows, however, 
that this results in a negative value. If we assume that $\alp=\Upsilon\alpq$, then there are positive $\kappa$'s for 
$\Upsilon$'s relatively close to $0.5$, as given in Table 1. Note that $\kappa=\frac{1}{4}$ corresponds to 
\begin{table}[htbp]
\centering
\begin{tabular}{l l l l l l l}
\vspace{.2cm} $\Upsilon$\quad &0.50\quad\quad &0.45\quad\quad & 0.42\quad\quad & 0.38\quad\quad & 0.27\quad\quad 
\\
\vspace{.2cm} $\kappa$\quad &\hspace{-0.12cm}-0.11\quad\quad &\hspace{-0.12cm}-0.03\quad\quad & 0.02\quad\quad & 
0.07 \quad\quad & 0.25\quad\quad
\end{tabular}
\caption{Estimates of the parameter $\kappa$.}
\end{table}
$\Upsilon=0.27$ versus $\Upsilon\approx 0.5$  in lattice QCD. It is possible to partially fix this discrepancy by choosing smaller $\kappa$. Then 
the value one needs is bounded from above by a number of order $0.02$. 

The problem is, of course, that the action for the baryon vertex we used in \eqref{potential} is oversimplified. There are extra background 
scalars as well as $\alpha'$ corrections.\footnote{In contrast to AdS/CFT, where $\frac{\alpha'}{R^2}\sim\frac{1}{\sqrt{\lambda}}$ becomes 
small for large 't Hooft coupling, the phenomenological estimate of \cite{az1} gives $\frac{\alpha'}{R^2}\sim 1$. The latter 
suggests that $\alpha'$ corrections are relevant for the real world and a departure from the supergravity approximation is needed.} However, 
there is also a possibility that the string formulation being reliable at large distances provides only a qualitative description of the physics at 
short distances.

\subsection{Y-Law and String Tension}
Apparently, one of the requirements for the multi-quark model should be consistency with the quark-antiquark case: as long as the fundamental 
strings are prevented from getting deeper into the $r$-direction, the baryon vertex, as a string endpoint, should also be prevented from doing so. Our 
next goal will be to understand the large distance behavior of the multi-quark potential from this point of view.

To implement this approach, we assume that: 1) the baryon potential ${\cal V}(\r0 )$ is a positive and regular function of $\r0$ such that it reaches 
the minimum exactly at $\r0=1/\sqrt{\s}$; 2) ${\cal V}\rightarrow +\infty$ as $\r0$ tends to zero or infinity. 
The latter means that the vertex can't come very close to the boundary as well as go far away from it. 

These conditions allow us to conclude that the last equation of \eqref{xyr-eqs} makes sense only if $\r0$ is 
subject to the constraint \eqref{wall2}. We also learn that $k_i\rightarrow 0$ as $\lambda\rightarrow 1$.  

If we look at Eq.\eqref{l-eqs}, then a short inspection shows that $l_i$ takes large values only in the corner of the parameter space $(k_i,\lambda)$ 
located at $(0,1)$. In this region $l_i$ behaves as

\begin{equation}\label{Largeli}
l_i=-\oh\sqrt{\frac{1}{\s}}\ln\bigl(\sqrt{k_i}+1-\lambda\bigr)+O(1)
\,.
\end{equation}
Note that in the symmetric case $k_i\sim (1-\lambda)^2$. Hence, the equation reduces to \eqref{largelE}.

A precisely analogous computation for $E$ gives

\begin{equation}\label{LargeE}
E=-\oh\ep\g\sqrt{\s}\sum_{i=1}^3\ln\bigl(\sqrt{k_i}+1-\lambda\bigr)
+O(1)
\,.
\end{equation} 
A contribution from the baryon vertex is of order $1$, because the function ${\cal V}$ is finite at $\lambda=1$. Combining this 
with \eqref{Largeli}, we find the energy configuration as a function of the lengths $l_i$

\begin{equation}\label{LargeE2}
E=\sig \sum_{i=1}^3l_i+O(1)
\,.
\end{equation}
Here $\sig$ is equal to $\tens$ of \cite{az1}, as expected from the universality of the string picture.

However, this is not the whole story. We still need to show that $Y$ is the Fermat point of the triangle formed by the quarks. The simplest way to 
see this is to take the limit $k_i\rightarrow 0$ in the first two equations of \eqref{xyr-eqs} which determine the location of the vertex on 
the $x\text{-}y$ plane. We have thus

\begin{equation}\label{Fermat}
\sum_{i=1}^3 \frac{1}{l_i}(\x0 -x_i)=0
\,,
\qquad
\sum_{i=1}^3 \frac{1}{l_i}(\y0 -y_i)=0
\,.
\end{equation}
Obviously, these equations is a result of differentiating the sum $\sum_{i=1}^3 l_i$ with respect to $\x0$ and $\y0$. Because a solution 

\begin{equation}\label{solution}
\x0=\frac{1}{{\cal N}}\sum_{i=1}^3 \frac{x_i}{l_i}
\,,\qquad
\y0=\frac{1}{{\cal N}}\sum_{i=1}^3 \frac{y_i}{l_i}
\,,\qquad
{\cal N}=\sum_{i=1}^3\frac{1}{l_i}
\end{equation}
determines the Fermat point, in this limit $Y$ tends to the Fermat point of the triangle. This completes our derivation of the Y-law. 

\subsection{The Case $SU(N)$}

Now we consider the case $SU(N)$. The analysis is a little bit more complicated but the results are very similar.

By analogy with section 2.1, we place $N$ heavy quarks on the boundary of the five-dimensional space. Because in a generic case 
the quarks are not on the same plane, we need to involve the third spatial coordinate $z$. So, the quarks are fixed in $\text{\bf R}^3$ at 
points $Q_i$ with coordinates $x_i,\,y_i,\,z_i$. The configuration of interest is constructed as the $N$ quarks joined together by 
the $N$ fundamental strings ending at the baryon vertex in the interior of the five-dimensional space whose metric takes the 
form \eqref{metric}. 

The total action of the system is given by \eqref{stringaction}, where the sum now runs from $1$ to $N$. After choosing the gauge \eqref{gauge}, 
the action of the $i$-string takes the form

\begin{equation}\label{N-ng-action}
S_i=T\g\int_0^1 d\sigma_i\,w\sqrt{a^2_i+ x'^2_i+z'^2_i+r'^2_i}
\,.
\end{equation}
The boundary conditions are given by Eq.\eqref{boundary} together with 

\begin{equation}\label{Nboundary}
z_i(0)=z_i\,,
\qquad
z_i(1)=\z0
\,.
\end{equation}

In addition to the first integrals of section 2.1, which correspond to Euler-Lagrange equations for the fields $x_i(\sigma_i)$ and $r_i(\sigma_i)$, 
there is one more integral due to translational invariance in the $z$-direction. Combining it with 
\eqref{1integral} results in 

\begin{equation}\label{N-2integral}
\bar P_i=z'_i\,.
\end{equation}
These equations yield solutions

\begin{equation}\label{z}
z_i (\sigma_i )=(\z0 -z_i)\sigma_i +z_i
\,
\end{equation}
which coincide at the endpoints with the boundary values defined in \eqref{Nboundary}.

Extremizing the action with respect to the location of the vertex, we get \eqref{xyr-eqs}, with the upper bound of summation given by $N$, and

\begin{equation}\label{pre-Fermat-Weber}
\sum_{i=1}^N\frac{\z0 -z_i}{l_i\sqrt{1+k_i}}=0
\,
\end{equation}
that determines the location of the baryon vertex along the $z$-direction. Note that now $l_i=\sqrt{(\x0 -x_i)^2+(\y0 -y_i)^2+(\z0 -z_i)^2}$.

Like in the case $N=3$, we can compute the expressions for the lengths $l_i$. After doing so, we find that they take the forms as in \eqref{l-eqs} 
and \eqref{energy3}, with the only difference: $E$ now includes the contributions from the $N$ fundamental strings.

It is also straightforward to extend the analysis of section 2.3 to the case of interest. Assuming the same form of ${\cal V}$, we quickly come to 
the similar conclusions about the bound \eqref{wall} and large values of $l_i$. Continuing along those lines leads to the asymptotic behaviors as 
\eqref{Largeli} and \eqref{LargeE}. Finally, we get  

\begin{equation}\label{LargeEN}
E=\sigN \sum_{i=1}^Nl_i+O(1)
\,.
\end{equation}
As before the string tension $\sigN$ is equal to that of the quark-antiquark case.

To complete the picture, we should check that $E$ given by the above formula is minimum. The essential point is to take the limit 
$k_i\rightarrow 0$ in the equations determining the vertex location. Then, from Eqs.\eqref{Fermat} and 

\begin{equation}\label{Fermat-Weber}
\sum_{i=1}^N\frac{1}{l_i}(\z0 -z_i)=0
\,,
\end{equation}
we learn that $Y$ is nothing but the geometric median of the set of the points $Q_i$.\footnote{It is also known as the Fermat-Weber 
point. See, e.g., \cite{fermat-weber}.} This is the desired result which reflects the fact that $E$ is minimum.

\section{Calculating the Pseudo-Potentials}
\renewcommand{\theequation}{3.\arabic{equation}}
\setcounter{equation}{0}
In this section we will investigate the temperature dependence of the spatial string tension. The model of interest is developed 
for the quark-antiquark case in \cite{az2,andreev}, to which the reader is referred for more detail. The philosophy is that the spatial 
string tension is determined from temperature dependent pseudo-potentials extracted from the spatial Wilson loops.\footnote{These are 
Wilson loops in hyperplanes orthogonal to the temporal direction.}

\subsection{General Formalism}

As before, we begin with the case $SU(3)$. Let $\WS$ be a spatial Wilson loop in ${\bf R}^3$, at a fixed value of $t$, with 
the path-ordered exponents $U_i$ along the lines shown in Figure 5.

%
\begin{figure}[ht]
\begin{center}
\includegraphics[width=2.5cm]{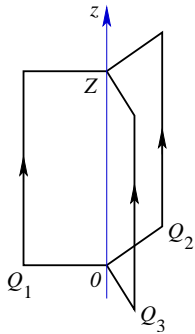}
\caption{\small{A baryonic spatial Wilson loop $\W$. The points $Q_i$ are on the $x$-$y$ plane.}}
\end{center}
\end{figure}

In the limit $Z\rightarrow\infty$ the expectation value of the loop is given by 

\begin{equation}\label{wloopS}
\langle\,\WS \,\rangle\sim \ep^{-Z\potS}
\,,
\end{equation}
where $\potS$ is a pseudo-potential.

Following \cite{az2,andreev}, we take the following ansatz for the five-dimensional geometry which turned out to be quite 
successful in the quark-antiquark case

\begin{equation}\label{metricT}
ds^2=R^2 w \bigl(f dt^2+d\vec x^2+f^{-1}dr^2\bigr)\,,
\qquad
f=1-\frac{r^4}{\rt^4}
\,,
\end{equation}
where $\rt=1/\pi T$. Note that it is nothing but a deformation of the  black hole in $\text{AdS}_5$.

To compute the spatial loop in question, we study a configuration similar to that shown in Figure 2: three fundamental strings, with the same 
orientation, such that the $i$-string begins at $Q_i$ and ends on the baryon vertex in the interior. The total action of the system is then given by 
the sum of the string actions and the action for the vertex. Because we are interested in the configuration independent of $z$, for the worldsheet 
coordinates we can choose

\begin{equation}\label{gaugeS}
z_i(\tau_i)=\tau_i\,,\quad
y_i(\sigma_i)=a_i\sigma_i+b_i
\,.
\end{equation}
With such a choice, the action of the $i$-string is then

\begin{equation}\label{ng-actionS}
S_i=Z\g\int_0^1 d\sigma_i\,w\sqrt{a^2_i+ x'^2_i+f^{-1}r'^2_i}
\,.
\end{equation}
As before, for the baryon vertex we take the action of the form

\begin{equation}\label{vertexS}
S_{\text{\tiny vert.}}=Z{\cal V}(\r0 )
\,.
\end{equation}
The boundary conditions on the fields are given by \eqref{boundary}. These determine the coefficients $a_i$ and $b_i$ in \eqref{gaugeS}.

To find the configuration, we must extremize the total action with respect to the worldsheet fields $x_i(\sigma_i)$ and $r_i(\sigma_i)$ as 
well as with respect to the coordinates of the vertex $\x0$, $\y0$, $\r0$. The underlying symmetries of the model provide a couple of first integrals 
that simplifies the analysis. Since the integrand in \eqref{ng-actionS} does not depend explicitly on $\sigma_i$, we have an integral

\begin{equation}\label{1integralS}
J_i=\frac{w_i}{\sqrt{a_i^2+x_i'^2+f^{-1}r_i'^2}}
\,.
\end{equation}
Like in section 2.1, translational invariance along the $x$-direction yields another integral. Combining it with \eqref{1integralS} results 
in \eqref{2integral}. Once this integral has been found, the $x_i$'s can be determined. As a result, we obtain \eqref{x}.

Now, we extremize the action with respect to the location of the baryon vertex. At finite temperature, the equations \eqref{xyr-eqs} are 
changed as follows. There is a factor $f^{-1}$ in front of $k_i$. Explicitly, 

\begin{equation}\label{xyr-S}
\sum_{i=1}^3 \frac{\x0 -x_i}{l_i\sqrt{1+\f0^{-1}k_i}}=0
\,,
\quad
\sum_{i=1}^3 \frac{\y0 -y_i}{l_i\sqrt{1+\f0^{-1}k_i}}=0
\,,
\quad
\sum_{i=1}^3\frac{1}{\sqrt{\f0^{-1}+k_i^{-1}}}+
\frac{1}{\g}\frac{\f0}{w}{\cal V}'(\r0)=0
\,,
\end{equation}
where $\f0 =1-\frac{\r0^4}{\rt^4}$. By virtue of \eqref{1integralS}, the integral over $\left [0,1\right]$ of $\sigma_i$ is equal to 

\begin{equation}\label{li-S}
l_i=\sqrt{\frac{\lambda}{\s(1+\f0^{-1}k_i)}}
\int_0^1 dv_i\,v_i^2\ep^{\lambda(1-v_i^2)}
\biggl(1-\Bigl(\frac{\lambda}{s\rt^2}\Bigr)^2v_i^4\biggr)^{-\oh}
\biggl(1-\frac{1}{1+\f0^{-1}k_i}v_i^4\ep^{2\lambda (1-v_i^2)}\biggr)^{-\oh}
\,.
\end{equation}
Here we have expressed the integration constant via the values of the fields at $\sigma_i=1$.

Now we will present an expression for the energy (pseudo-potential) of the configuration. The computation is just 
as above.\footnote{ Indeed, our previous discussion in section 2.1 corresponds to the case $f=1$.} At the end of the day, we have

\begin{equation}\label{E-S}
\tilde E={\cal V}(\lambda)
+\g\sqrt{\frac{\s}{\lambda}}\sum_{i=1}^3
\int_0^1 \frac{dv_i}{v_i^2}\Biggl[
\ep^{\lambda v_i^2}
\biggl(1-\Bigl(\frac{\lambda}{\s\rt^2}\Bigr)^2v_i^4\biggr)^{-\oh}
\biggl(1-\frac{1}{1+\f0^{-1}k_i}v_i^4\ep^{2\lambda (1-v_i^2)}\biggr)^{-\oh}-1-v_i^2\Biggr] + C
\,,
\end{equation}
where $C$ is a normalization constant. As a result, the pseudo-potential is given in parametric form by Eqs.\eqref{xyr-S}-\eqref{E-S}.

\subsection{Y-law and Spatial String Tension}

Now we consider the analog of the Y-law for the pseudo-potentials. For this, we continue to assume that ${\cal V}(\r0 )$ is a positive and regular 
function with a minimum at $\r0 =1/\sqrt{\s }$ and singularities at the endpoints. 

For this choice of ${\cal V}$, we can now determine the allowed values of $\r0$. A simple analysis shows that the 
equations \eqref{xyr-S} are consistent if $\r0<\min\left(1/\sqrt{\s},\rt \right)$ or $\r0>\max\left(1/\sqrt{\s},\rt\sqrt[4]{1+k_i}\right)$. Since 
the latter yields complex values of $l_i$, it must be omitted. We have thus

\begin{equation}\label{wallS}
\r0 \leq\min \left(1/\sqrt{\s},\rt \right)
\,.
\end{equation}
What we have found is nothing but the bound of \cite{az2,andreev}. In addition to \eqref{wall2}, it now states that neither strings nor baryon 
vertices may go behind the horizon $(r=\rt )$. From the point of view \cite{az2}, the bound gives rise to the two walls: 1) If $1/\sqrt{\s }<\rt$, then 
the first wall, $\r0=1/\sqrt{\s}$, dominates. This is interpreted as the low temperature phase; 2) If $\rt <1/\sqrt{\s}$, then the second wall, 
$\r0=\rt$, dominates. This is the high temperature phase. The phase transition point is at $\rt=1/\sqrt{\s}$. In terms of a critical temperature and 
the parameter $\s$, it is written as 
\begin{equation}\label{Tc}
\T=\frac{1}{\pi}\sqrt{\s}
\,.
\end{equation}

Having determined the allowed region for the parameter $\r0$, we are now ready to study $l_i$ as a function of two variables. To this end, we look 
for level curves in the allowed region.\footnote{As an illustration, we take $k_i\leq 2$. We do not elaborate on upper bounds for the $k_i$'s, 
because it is irrelevant for our purposes. } A summary of our numerical results is shown in Figure 6. The first observation is that $l_i$ takes large 
values 
\begin{figure}[htp]
\centering
\includegraphics[width=4cm]{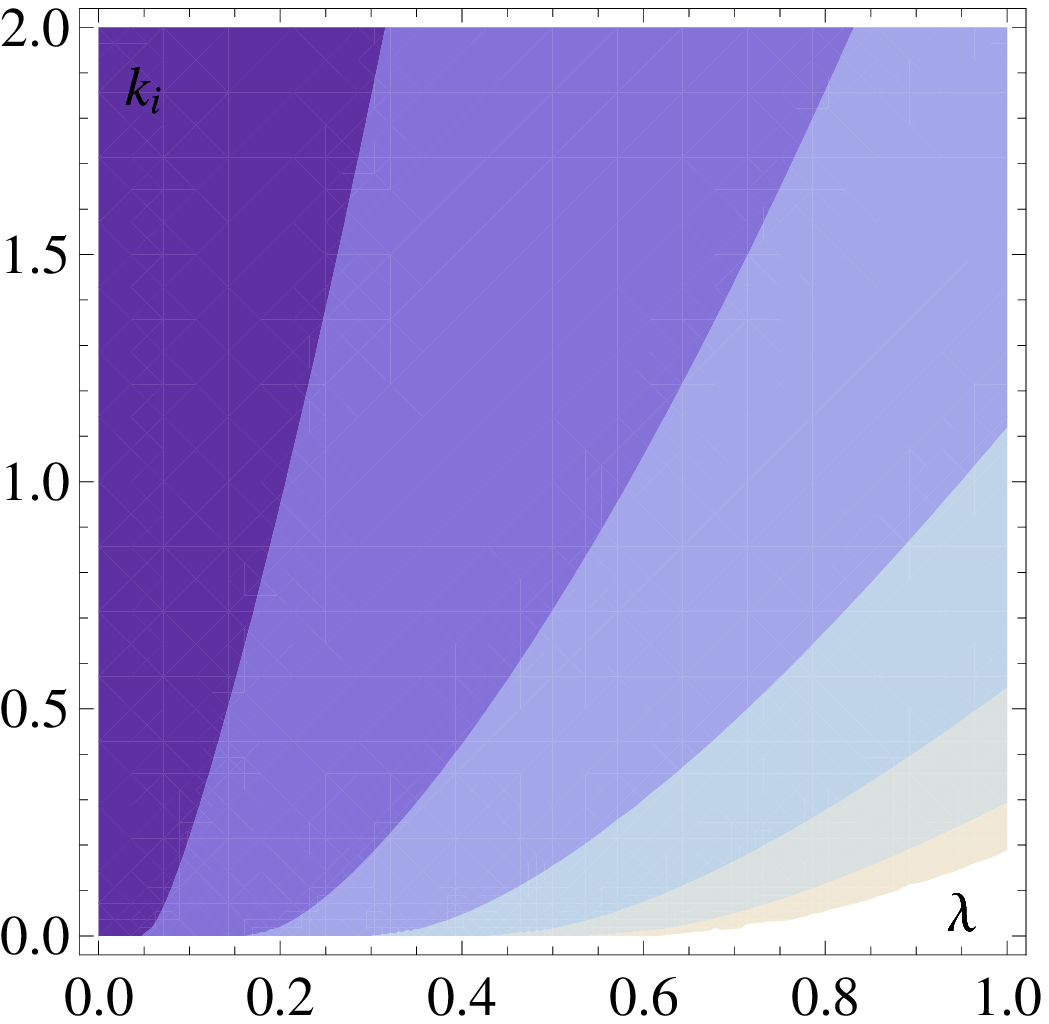}
\hspace{3cm}
\includegraphics[width=4cm]{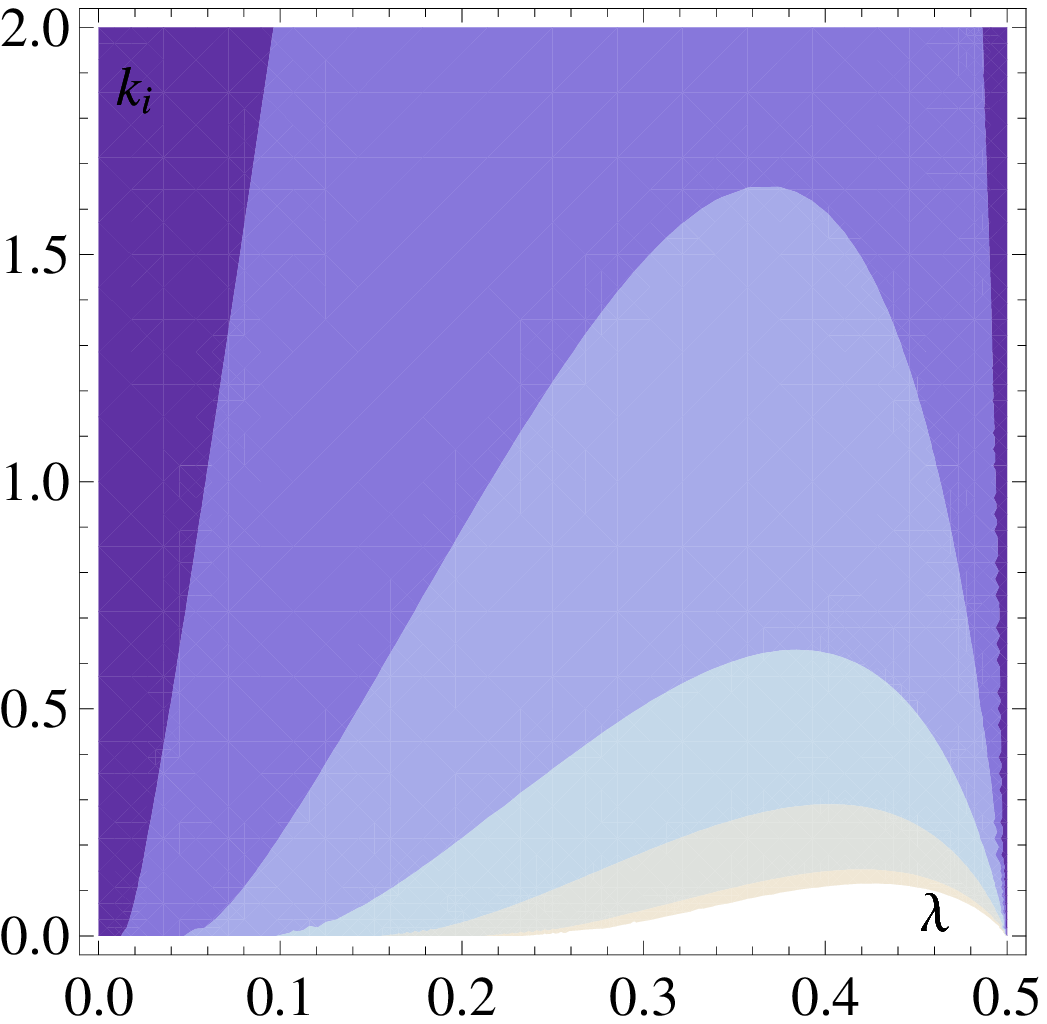}
\\
 \caption{\small{Level curves for $l_i=r(\lambda ,k_i)$: in the low temperature phase (left) at $T=\T/\sqrt{2}$ and in 
the high temperature phase (right) at $T=\sqrt{2}\T$. Larger values are shown lighter. Here $\s=0.45\,\text{GeV}^2.$}}
\end{figure}
only in the vicinity of the lower right corner which lies on the $\lambda$ axis at $\lambda=1$ if $T<\T$ and at $\lambda=\s\rt^2$ if $T>\T$. 
Another important point is that at nonzero $k_i$ the length $l_i$ vanishes at $f=0$. This fact implies that in the high temperature 
phase we should keep $k_i/\f0 \rightarrow 0$ to get large lengths.

According to the numerical analysis, $l_i$ may become large only if $k_i$ is small and $\r0$ saturates the bound \eqref{wallS}. It follows from 
\eqref{li-S} and \eqref{E-S}, that in this limit the length and the energy behave as 

\begin{align}\label{largeli-S}
l_i&=-\frac{1}{2}\sqrt{\frac{\lambda}{\s}}
\ln\Bigl(\beta\sqrt{k_i}+k_i\f0^{-1}(1-\f0)+\f0 (1-\lambda)\Bigr)^{\frac{1}{\beta}}+O(1)
\,,\\
\tilde E&=-\oh\g\ep^\lambda\sqrt{\frac{\s}{\lambda}}
\sum_{i=1}^3\ln\Bigl(\beta\sqrt{k_i}+k_i\f0^{-1}(1-\f0 )+\f0 (1-\lambda)\Bigr)^{\frac{1}{\beta}}
+O(1)
\,,
\end{align}
where $\beta^2=4-\frac{3}{2}k_i\f0^{-1}+\frac{3}{2}k_i-\frac{11}{2}\f0+\oh\lambda (17\f0-8)-2\lambda^2 \f0$. From these formulas, it is 
evident that at large distances the pseudo-potential is given by 
\begin{equation}\label{largeE-S2}
\tilde E=\sigS\sum_{i=1}^3l_i+O(1)
\,,
\end{equation}
with the spatial tension
\begin{equation}\label{tension-s}
\sigS=
\begin{cases}
\sig & \text{if} \quad T\leq\T\,,\\
\sig \bigl(\tfrac{T}{\T}\bigr)^2 \exp\bigl\lbrace \bigl(\tfrac{\T}{T}\bigr)^2-1\bigr\rbrace 
& \text{if} \quad T\geq\T\,.
\end{cases}
 \end{equation}
Here, $\sig$ is the physical tension at zero temperature. The temperature dependence of $\sigS$ is the same as that of \cite{az2,andreev} found in 
the quark-antiquark case. This suggests that the spatial string tension is also universal. 

Finally, taking the limit $k_i\rightarrow 0$ in the first two equations of \eqref{xyr-S}, we learn that $Y$ is the Fermat point of the 
triangle $Q_1Q_2Q_3$. Thus, we have derived, in the case of pseudo-potential, the $Y$-law analogous to that of the quark potentials.

\subsection{Further Remarks}

We have here considered the five-dimensional framework for studying the spatial string tension. However, if one can describe the string theory 
construction, this gives the baryon vertex as a wrapped fivebrane. In this case, the brane world-volume is ${\bf R}\times{\bf X}$, with 
${\bf R}$ a {\it spacelike} curve in the deformed $\text{AdS}_5$.

It is worth noting that the temperature dependence of the spatial tension is in good agreement with the lattice data for 
$T\leq 3\T$ \cite{az2, andreev}. Note that at higher temperatures it is determined by the $\beta$-function of a pure $SU(3)$ gauge theory
\cite{stension-su3}. Clearly, our model cannot reproduce logarithms associated with the running coupling. Instead, it provides a complementary 
description in the strong coupling regime.

The above analysis can be easily generalized to the case $SU(N)$. Compared to the discussion of section 2.4, the points $Q_i$ are now on the $x$-$y$ 
plane. Hence, we needn't to involve any additional coordinate. The formal modification to be made is to extend the upper bound of summation to $N$. 
Thus, the temperature dependence of the spatial string tension is given by \eqref{tension-s} that is the desired result \cite{andreev}.

\section{Concluding Comments}
\renewcommand{\theequation}{4.\arabic{equation}}
\setcounter{equation}{0}
We have so far discussed the configurations including a single baryon vertex. In general, it is possible to consider configurations with an 
arbitrary number of vertices: string networks.

To give an idea of how this works, let us consider, though only schematically, the tetraquark case that is the four-quark potential $\tetra$. We place 
quarks and anti-quarks at boundary points $Q_i$ and $\bar Q_i$, respectively. These points are endpoints of fundamental strings. There are two 
basic configurations, as sketched in Figure 7. The connected configuration includes five strings beginning at the boundary points and 
%
\begin{figure}[ht]
\begin{center}
\includegraphics[width=9.25cm]{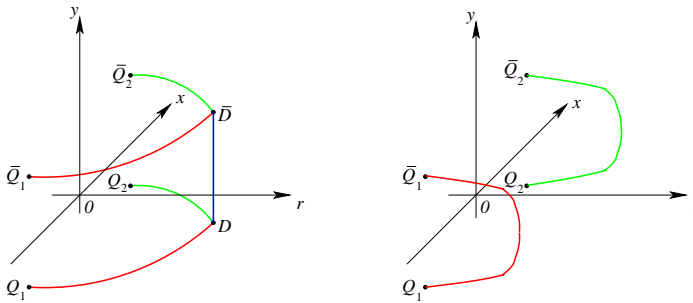}
\caption{\small{Configurations used to calculate the tetraquark potential.}}
\end{center}
\end{figure}
ending on the baryon and anti-baryon vertices in the interior of the five-dimensional space. The disconnected configuration, a "two-meson'' state, 
includes two strings stretched between the quarks and anti-quarks. 

For the tetraquark potential, there is a well-known "flip-flop''\cite{tetra}. When the quarks and anti-quarks are well separated, the potential is given 
by Coulomb terms plus a term which indicates the formation of the two-Y flux. On the other hand, when the quarks and anti-quarks are close, the 
potential becomes a sum of two quark-antiquark potentials. Clearly, these two possibilities correspond to our connected and disconnected 
configurations. 

From the point of view of 10-dimensional string theory, it is very easy to understand why the disconnected configuration dominates at small 
separations. The connected configuration includes the baryon and anti-baryon vertices  that is nothing but a 
brane-antibrane system in $10$ dimensions. There is a critical separation for such a system. For smaller separations, the open string spectrum 
consists of a tachyon mode. The instability associated with the tachyon represents a flow toward annihilation of the brane-antibrane system. 

\vspace{.25cm}
{\bf Acknowledgments}

\vspace{.25cm}
\noindent This work was supported in part by DFG "Excellence Cluster'' and the Alexander von Humboldt Foundation under
Grant No. PHYS0167. The author thanks G. Lopes Cardoso, D. Sorokin, V.I. Zakharov, and P. Weisz for discussions and comments.

\vspace{.35cm} 
{\bf Appendix}
\renewcommand{\theequation}{A.\arabic{equation}}
\setcounter{equation}{0}

\vspace{.25cm} 
\noindent The purpose of this appendix is to consider the two limiting cases of the equations \eqref{l-eqs-ex} and \eqref{energy-ex}: short and long 
distances.

We begin with the case of long distances. As in Figure 3, large $l$'s correspond to values of $\lambda$ near $1$. For $\lambda\rightarrow 1$, 
the integrals in \eqref{l-eqs-ex} and \eqref{energy-ex} can be evaluated to give \footnote{The integrals are dominated by $v\sim 1$, where they 
take the form $\int  dv/\sqrt{a+b(1-v)+c(1-v)^2}$. The remaining integral may be found in \cite{ryz}.}

\begin{equation}\label{largelE}
l=-\oh\sqrt{\frac{1}{\s}}\ln\left(1-\lambda\right)+O(1)
\,,
\qquad
E=-\frac{3}{2}\ep\g\sqrt{\s}\ln\left(1-\lambda\right)+O(1)
\,.
\end{equation}
Now, the parameter $\lambda$ is easily eliminated, and one gets 

\begin{equation}\label{Y-law}
E=\sig\lmin +O(1)
\,,
\end{equation}
where $\lmin=3l$ and $\sig=\ep\g\s$. 

In a similar spirit, we can explore the short distance behavior of $E$. For $\kappa\leq 1$, this is simple because we need to take the 
limit $\lambda\rightarrow 0$ (see Fig.3). Expanding the right-hand sides of \eqref{l-eqs-ex} and 
\eqref{energy-ex} in powers of $\lambda$, at leading order we obtain 

\begin{equation*}\label{lE}
l=\lambda^\oh\bigl(l_0+O(\lambda)\bigr)
\,,
\qquad
E=-\lambda^{-\oh}\bigl(E_0 +O(\lambda^\oh)\bigr)
\,,
\end{equation*}
where $l_0=\sqrt{\tfrac{1-\kappa^2}{9\s}}\,{}_2F_1\bigl[\tfrac{1}{2},\tfrac{3}{4};\tfrac{7}{4};1-\kappa^2\bigr]$ and 
$E_0=3\g\sqrt{\s}\bigl({}_2F_1\bigl[-\tfrac{1}{4},\tfrac{1}{2};\tfrac{3}{4};1-\kappa^2\bigr]-\kappa\bigr)$. Combining these formulas, 
we find

\begin{equation}\label{Coulomb}
E=-3\frac{\alp}{L}+O(1)
\,,
\end{equation}
where $\alp=l_0E_0/\sqrt{3}$.

The case $\kappa=1$ can be treated similarly. Our previous formulas  for $l$ and $E$ become

\begin{equation*}\label{El=1}
l=l_1\lambda+O(\lambda^2)
\,,
\qquad
E=C+E_1\lambda^\oh+O\bigl(\lambda^{\frac{3}{2}}\bigr)
\,,
\end{equation*}
where $l_1=1/\sqrt{3\s}$ and $E_1=6\g\sqrt{\s}$. This implies that

\begin{equation}\label{Er=1}
E=C+\gamma L^\oh +O(L)
\,,
\end{equation}
where $\gamma=E_1/\sqrt[4]{3l_1^2}$.

For larger values of $\kappa$, the analysis is a bit more involved because of the lower bound \eqref{wall2}. In this case we need to take the 
limit $\lambda\rightarrow\last$ (see Fig.3). Note that $\rho(\lambda_\ast)=0$. A simple but somewhat tedious calculation shows that in the 
neighbor of $\lambda=\last$ the length $l$ and the energy behave as

\begin{equation*}\label{l>1}
l=(\lambda-\last)^\oh\bigl(l_0^\ast+O(\lambda-\last )\bigr)
\,,
\qquad
E=E_0^\ast+E_1^\ast (\lambda-\last)+O\bigl((\lambda-\last )^2\bigr)
\,,
\end{equation*}
with  
\begin{equation*}
l_0^\ast=\frac{1}{4}\sqrt{\frac{\rho'}{\s\last}}
\biggl(\sqrt{\frac{\pi}{\last}}\text{erf} (\sqrt{\last} )-2\biggr)
\,,
\,\,\,
E_0^\ast=E(\last )
\,,
\,\,\,
E_1^\ast =\frac{3}{8}\g
\rho'
\ep^{2\last}
\sqrt{\frac{\s}{\last^3}}
\biggl(\sqrt{\frac{\pi}{\last}}\text{erf} (\sqrt{\last} )-2\ep^{-\last}\biggr)
\,. 
\end{equation*}
Here $\rho'$ denotes the derivative of $\rho$ at $\lambda=\last$. This implies that at short distances the potential is given by 

\begin{equation}\label{Er>1}
E=E_0^\ast+\gamma L^2 +O(L^3)
\,,
\end{equation}
where $\gamma=E_1^\ast/3(l_0^\ast)^2$.


\end{document}